\RequirePackage{fix-cm}
\documentclass[twocolumn]{svjour3}          
\smartqed  
\usepackage{graphicx}
\begin{document}

\title{
The archive solution for distributed workflow management agents of the CMS experiment at LHC.
}


\author{
Valentin Kuznetsov \and
Nils Leif Fischer \and
Yuyi Guo
}


\institute{Valentin Kuznetsov \at
              Cornell University, USA \\
              \email{vkuznet@gmail.com}
           \and
           Nils Leif Fischer \at
              Heidelberg University, Germany \\
              \email{n.fischer@stud.uni-heidelberg.de}
           \and
           Yuyi Guo \at Fermilab National Laboratory, USA \\
           \email{yuyi@fnal.gov}
}

\date{Received: date / Accepted: date}

\maketitle

\begin{abstract}
The CMS experiment at the CERN LHC developed the Workflow Management
Archive system to persistently store unstructured framework job
report documents produced by distributed workflow management agents. In
this paper we present its architecture, implementation, deployment, and
integration with the CMS and CERN computing infrastructures, such as central HDFS
and Hadoop Spark cluster. The system leverages modern technologies
such as a document oriented database and the Hadoop eco-system to provide the necessary
flexibility to reliably process, store, and aggregate $\mathcal{O}$(1M) documents on a daily basis.
We describe the data transformation, the short and long term storage layers,
the query language, along with the aggregation pipeline developed to visualize
various performance metrics to assist CMS data operators in assessing the
performance of the CMS computing system.

\end{abstract}

\section{Introduction}\label{sec:Introduction}
The success of physics programs at CERN would be impossible without reliable
data management systems. During the Run 1 and Run 2 data taking periods
four experiments at the Large
Hadron Collider (LHC) in Geneva, Switzerland, demonstrated the ability to handle petabytes of data on a regular
basis. In CMS, we rely on a tiered computing infrastructure.
The data collected with the detector are streamed to the high level trigger (HLT)
farm and organized into trigger streams. Later they are archived at the Tier-0
center at CERN and distributed to CMS Analysis facilities at CERN
and Tier-1 centers around the globe. Many Tier-2 centers worldwide
share a portion of the data for further processing and Monte Carlo (MC) generation.
Finally, the Tier-3 centers (mostly at Universities) are used for various
analysis tasks. More details about CMS Computing Model can be found
elsewhere \cite{CMSComp}.

To accomplish various processing tasks, both real and MC data
are handled by a distributed set of agents at various computing centers.
The chain of high energy physics (HEP) workflows
is performed within an individual agent.
For instance, a job can run an MC
simulation where the CMS software (CMSSW) framework generates new data in a
distributed fashion, collects various pieces of data frame the workflow pipeline, and publishes
results in the central Data Bookkeeping System (DBS). At every step, the associated
metadata information are collected in ``framework job report'' (FWJR)
documents. This information can be further analyzed and used to
monitor the entire process both in terms of various metrics, e.g. successfull job throughput,
or capture various errors in different steps of the workflow pipeline. In addition, the information
can be used to plan future jobs and  better manage the resources.
Initially, ad-hoc solutions were used by various data-operations teams, but
it soon became clear it was necessary to unify the various solutions under a single system.

Here, we present the Worklfow Management Archive system (WMArchive), which was
designed to provide a long term solution for FWJR document storage along with
flexible queries and visualization to help data ops in their
daily operations.

This paper is organized as follows. In Sec. \ref{sec:Architecture}
we present the overall architecture of the system starting by
describing the CMS workflow management system (Sec.~\ref{sec:WMAgents}),
followed by discussion of the requirements (Sec.~\ref{sec:Requirements}).
We then present the system components, storage layers, and
interfaces in Secs.~\ref{sec:Components}, \ref{sec:Storage}, and \ref{sec:Interfaces},
respectively. In Sec.~\ref{sec:Benchmarks} we show various benchmarks
and discuss query language used in the WMArchive in Sec.~\ref{sec:QL}. We finish our
discussion with current use of the system in Sec.~\ref{sec:Monitoring}
and provide an overview of the monitoring tools based on stored metadata.

\section{WMArchive architecture}\label{sec:Architecture}
The WMArchive system architecture is quite complex.
Here, we start our discussion with
overview of the CMS workflow management system followed by
set of requirements imposed by our data operations teams.

\subsection{CMS Workflow Management System}\label{sec:WMAgents}

The CMS Workflow Management System~\cite{WMAgent} contains several applications to process the
workflows defined by physics groups.  The overview of the procedure is as
follows.  When a workflow is fed into the system (Request Manager), it divides
the work into smaller units which are placed into a queue (WorkQueue).  Individual
Workflow Manager Agents (WMAgents), distributed among computing centers,
pull down work from the queue depending on resource availability and other
various conditions, e.g. priority of the work, etc.  Each WMAgent is responsible
for splitting the work into smaller chunks (jobs) and sending them to 
the CMS Global Pool~\cite{GlobalPool}, an HTCondor~\cite{HTCondor} batch system overlayed, 
using GlideinWMS~\cite{GlideinWMS}, on the CMS-accessible grid resources.
The batch system then distributes the jobs to
computing resources all over the world---the tiered system mentioned above.  Each
WMAgent is also responsible for tracking and accounting for the jobs and
publishing the results in other applications in CMS (DBS, PhEDEx, WMArchive).
We have about a dozen WMAgents running concurrently. 

When a job has finished
sucessfully or failed, WMAgent will generate  an FWJR which 
contains job specific information, such as the input/out datasets and files,
CPU and memory usage, log file and its location, and so on.

The current set of WMAgents can produce on average 500K FWJRs per day but could not keep
the documents in the system indefinitely. FWJR reports are published and stored
in WMArchive for further analysis and as a solution for long term storage
needs.

\subsection{Requirements}\label{sec:Requirements}

In CMS we rely on the distributed nature of WMAgents. At the time of initial design of the system, we assumed
the WMAgents would generate 200--300\,000 FWJR documents per day, with a document size of 
$\mathcal{O}$(10~KB) which would yield 3~GB total/day or about 1--2~TB/year of
metadata information. 
Each FWJR document is in a JSON data format whose structure is
specific to the job. For example, the FWJR of a successful job will be different
from a failed job. Here, is a list of requirements we gathered from CMS data
management and workflow management teams:
\begin{itemize}
\item store FWJR for the lifetime of the experiment in persistent storage for
    data lookup anytime; losing data is not acceptable;
\item provide data-ops quick and flexible queries to retrieve the log files of
    any particular job for debugging;
\item the statistics to be collected include, but are not limited to, CMSSW version,
    CPU/wallclock and memory information, log files and their locations, input and
        output datasets, number of files and events, job error codes, etc.
\item the system should be able to use a flexible query language over
    unstructured data; there is no requirement to support live queries;
\item provide tools to query, aggregate, and analyze CMS jobs and to plan for
     future processing and resource management;
\item provide a web interface to monitor job processing trends, gather statistics,
    and aggregate information across multiple dimensions;
\item provide a web interface to hourly update job stats for debuggging and error
    reporting of running jobs.
\item support a flexible schema to extend, add, or remove fields from the FWJR if necessary;
\item have a minimal impact on existing CMS infrastructure, in particular on production WMAgents;
\end{itemize}

\subsection{System architecture}\label{sec:Components}

\begin{center}
\begin{figure}
\includegraphics[width=0.5\textwidth]{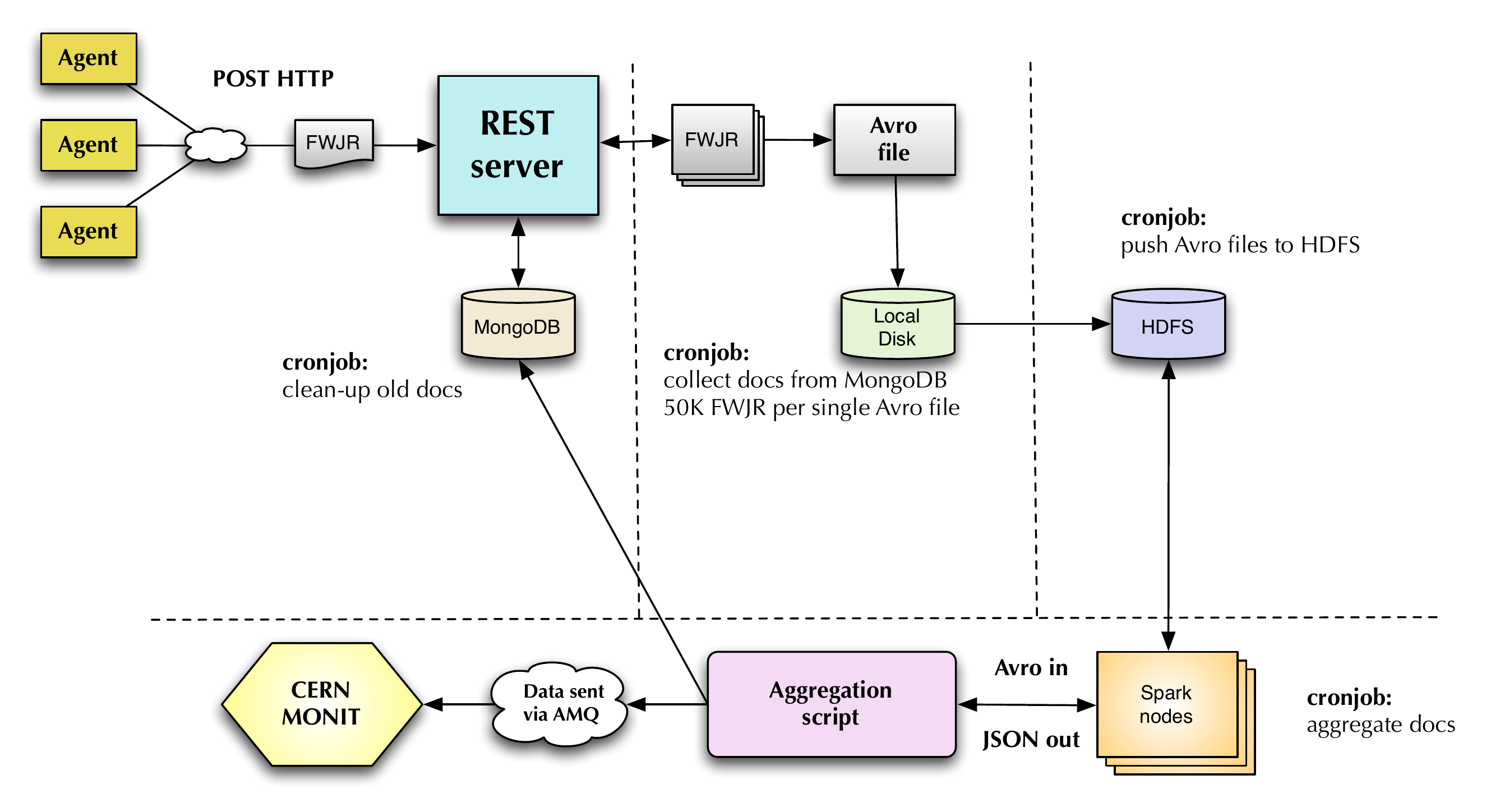}
    \caption{WMArchive architecture. MongoDB represents short-term storage (STS)
    and HDFS refers to long-term storage (LTS) discussed in a text.}
\label{fig:Architecture}
\end{figure}
\end{center}

WMArchive system is based on a RESTful server and series of tools to handle
various data management tasks. The overall architecture of the system is shown
in Fig. \ref{fig:Architecture}.  The data are pushed to WMArchive by the
distributed set of WMAgents via REST APIs. All WMAgents are authenticated via
CMSWeb by providing proxy GRID certificates at front-end servers. The
pushed data are submitted as a bulk collection in certain time intervals defined
by WMAgent configuration.\footnote{Right now every WMAgent sends data every 5 minutes, 1000
docs per polling cycle with rate of 200 docs per single HTTP request.}
The received data are routed to internal
short-term storage (STS) based on the document-oriented MongoDB database~\cite{MongoDB}.  
Then, a separate daemon reads data from STS, merges records
together, converts the JSON data to Avro~\cite{Avro} data format, and writes Avro files
to the local file system.  Finally, we rely on a migration script to push data from the
local file system into long-term storage (LTS) based on HDFS. Once data are
written to LTS we use an aggregation script to collect hourly and daily
statistics and push them into STS and the CERN MONIT systems.

\subsection{Storage layers}\label{sec:Storage}
Based on the system requirements (see discussion in Sec.~\ref{sec:Requirements}), we
decided to develop a set of abstract interfaces to provide access to available storage solutions.
Later, we provided a common implementation of BaseIO, FileIO, MongoIO, and
HdfsIO interfaces, which were used for rapid prototyping of APIs and
understanding interprocess communications.  After several trials (see 
Sec.~\ref{sec:Benchmarks}), we decided on the short and long term storage
solution described above. The former is responsible for keeping
up with the injection rate while the latter used for permanent storage of FWJR
documents. In other words, we use STS as a buffer to keep up with our agents'
load, while LTS serves as persistent storage.

Among the various candidates, we decided to use MongoDB with WiredTiger
document-level concurrency control as our STS solution. We estimated that our
injection rate should stay within 1 KHz and tested that MongoDB will sustain
such load. While we did not impose any constraint on the incoming document
structure, we complemented each document with auxiliary information such as
WMArchive unique id and storage type (e.g.  \emph{fileio}, \emph{mongoio}, \emph{Avroio})
attributes.  The latter was used by migration and clean-up scripts to
dynamically handle the size of STS within budget constraints.

For the LTS, we decided to use HDFS provided by the CERN IT infrastructure;
several issues we faced justifyied this choice.  First, our FWJR documents were
quite small (about 10~KB) to be stored individually on HDFS.
Therefore, we accumulated our data into larger groups before storing them to HDFS.
However, we also needed to satisfy another requirement (see Sec.~\ref{sec:Requirements}),
and provide flexible queries over a large portion of stored data.  As a result, we
decided to convert our data into the Avro~\cite{Avro} data format, which was a more
natural fit for both tasks. It guarantees  data consistency within
a single file and optimization for the HDFS eco-system such as efficient use of
Spark jobs to process large amounts of data.

We complemented the WMArchive service with three cron jobs, one, to continuously
accumulate data from STS to Avro files on local filesystem,  another   to move
Avro files from staging area to HDFS, and a third to clean-up STS. The first
was designed to collect enough documents to fit into the HDFS block boundaries.
The CERN HDFS system is using 256~MB block boundaries and was adjusted
by CERN IT to the needs of experiments based on different use-cases. This constrain
was added as one of the configuration parameters of WMArchive system.
Within this boundary we collected roughly 50\,000 documents
into a single Avro file.
The actual number of documents in Avro file varied based on a file
size growth up-to a given threshold (in our case 256~MB) rather than fixed
number of stored documents.
These files were then moved into a staged area where they
were periodically purged once relocated to HDFS by the second cronjob. Afterwards, 
the underlying script flipped the document's storage type attribute in STS from
\emph{mongoio} to \emph{Avroio}. These values of the attributes were used by the third cronjob which
performed a cleanup in STS of documents with the \emph{Avroio} storage type value based
on a predefined threshold. We determined this threshold based on the disk capacity
of the production system and the desire to keep documents in STS. In our case, we keep
documents in STS for 1 month after they are moved to LTS.

The data movement was optimized with respect to the accumulation rate of documents
into a single Avro file. Empirically, we found that a ten minute interval is
sufficient to accumulate 50\,000 (256~MB) docs into a single Avro file. Therefore, the
overall latency of data appearance on HDFS was approximately 10--20 minutes.
Minimization of this interval was important for the developement of monitoring
tools based on the aggregated information from LTS (see Sec.~\ref{sec:Monitoring}).

\subsection{Interfaces}\label{sec:Interfaces}

Users communicate with the WMArchive service via REST APIs. The POST APIs are
used both for data injection and placing queries into WMArchive. The latter
is done in real time in STS or via batch submission of Spark jobs on LTS.
To distinguish these use cases, we rely on a document representation rather than
individual APIs. For example, the data injection is done in a bulk using
the following document structure:
\begin{verbatim}
{"data":[list_of_documents]}
\end{verbatim}
while end-user queries are represented as:
\begin{verbatim}
{"spec":{conditions},
 "fields":[list_of_attributes_to_retrieve]} .
\end{verbatim}
These requests are acknowledged by the system which provides a response with a
unique token id to look-up their data at a later time, e.g.
\begin{verbatim}
{"job": {"results":{results_of_job},
         "wmaid":wma_unique_id}} .
\end{verbatim}
This approach allows us to process user-based queries on the HDFS/Spark cluster
independently from data injection tasks and other user communications with the system.
The results of the Spark jobs are stored back into STS and users are able to look-up
the results later via a GET request based on the provided token.
For example, to fetch the set of data from WMArchive end-users rely on the GET
API, e.g. 
\begin{verbatim}/wmarchive/data/wma_unique_id .\end{verbatim}

To minimize the execution time and to avoid a situation when user queries can
span the entire data collection (millions of documents per year), we require each
query to provide a time range attribute which specifies the boundaries of scanned
documents on HDFS. If both ranges of the time range are below a certain threshold
(implied by STS capacity) the query is executed in real time on STS,
otherwise a job submission is executed on LTS. The LTS job, i.e. Spark job on
HDFS, is routed to the CERN analytics cluster.  We performed various benchmarks to
determine latencies of queries executed on STS and LTS.

\section{WMArchive Benchmarks}\label{sec:Benchmarks}
To ensure that our python based code scales, we performed various benchmarks
using the STS and LTS back-ends.  The benchmarks were designed to test data
injection, parsing, and access patterns on a local node and a Spark cluster
provided by CERN IT. For local tests we used a Linux node with
24~GB of RAM and 12 cores; the CERN Spark cluster consists of 39 nodes with
64~GB of RAM and 32 cores per node.

\subsection{Short-Term Storage benchmark}\label{sec:MongoDBIndex}

CouchDB and MongoDB have been used by the group in different projects.
Initially, CouchDB was chosen to implement the WMArchive DB short-term storage
backend. However, it was not possible to have flexible queries 
nor the required volume of data storage. MongoDB was chosen for these reasons:
\begin{itemize}
\item it enables horizontal scaling of data;
\item it provides a flexible query language;
\item it is possible to exchange data beween the current CouchDB documents and MongoDB
    (WMAgents use CouchDB as temporate storage for FWJRs).
\end{itemize}

One disadvantage of MongoDB is that the query speed depends on indices which should fit
into the RAM of the node, for details see \cite{MongoDBIndex}.
If not, the lookup time increases significantly due to reading indices from the disk.
We studied our data in  MongoDB regarding the data size, index size, insertion speed, and query
response times.

We injected 1\,542\,513 real FWJR documents into MongoDB and created 13 indices on the
database. The total data size in the database was 15.3~GB and the total index size was 3.5~GB. 


Based on this profile we estimated that we should be able to hold about 10 M docs with our hardware and still have all the indices be able to fit
into the RAM (24~GB).  Found that  the  time  to insert 1000 documents
in bulk is less than 0.5 seconds. In addition, the insertion speed was
not affected by the database size. Uploading  data into MongoDB has a mininual
effect on WMAgent performance. Based on our studies, we concluded it was feasible to store one month of data
in MongoDB with the current hardware  and still maintain acceptable performance.

\subsection{Data lookup on a single node}
To test data access on a single node we used custom code to translate FWJR  documents 
from the JSON format into the Avro format. We generated
an Avro file with 50\,000 FWJR documents of identical structure. The total file size
in Avro format was about 190~MB. The bzip'ed Avro file shrank to 26~MB. Even
though we achieved such a large compression level, we decided to proceed with plain
Avro files on HDFS while rolling our system to production. The usage of zipped
Avro files was postponed to a later time to become more familiar with all
aspects of the system in the production environment.

We used Spark scripts to run a simple map-reduce job over the set of Avro
files. For this benchmark we decided to use a rate of 200\,000 documents/day, so we put four
Avro files into a single directory. Then we cloned that directory to simulate
two months of CMS data. In total we collected 12 million records. We ran Spark jobs and
measured the time we spent to find individual records based on provided logical
file name (LFN) patterns. We found that one day of data can be processed in
about a minute, while 2 months of data (12 million records) will require about 1 hour
of processing time.

\subsection{Data lookup on a Spark cluster}
After benchmarking data access patterns on a single node, we moved our tests to
the CERN IT Spark cluster. In addition to the 
search time we also measured the overhead between the Java and Python processes'
communication and found it satisfactory for our needs.

The tests were done in two modes:
\begin{itemize}
\item
    {\bf yarn-cluster:} all Spark execution actors were included inside the cluster
(driver and executors). This is the default setup provided by CERN IT.
\item
    {\bf yarn-client:} the Spark driver was spawned in the local node meanwhile executors
were inside the cluster. This setup was preferred for us since we were able to control
our production node utilization.
\end{itemize}

For both modes we used two months of data and placed a query to find a pattern
in task names across all documents stored on HDFS. The Spark jobs were configured to
use 4 executors, 4~GB of memory, and 4 cores each.

We obtained the following results: search results across one day of data was finished
in $\mathcal{O}(10)$ seconds, one month required $\mathcal{O}(100)$ seconds, and a search
pattern over two months of data fits into $\mathcal{O}(200)$ seconds.

\section{WMArchive Query Language}\label{sec:QL}
Our users, the CMS data-operations teams, require a flexible query language
(QL) to communicate with the WMArchive system. Since we didn't base our system on an RDBMS solution nor
impose live-query requirements, we opted  in favor of a JSON based QL
similar to the MongoDB QL \cite{MongoQL}. Due to the nested structure of stored metadata
within FWJR documents we used dot notations, e.g.  \emph{output.inputDataset},
to specify query conditions. We also allow users to
use flexible conditions such as $\$gt$, $\$lt$ operators as well as use an alternative set
of conditions via $${"\$or": [JSON1, JSON2]}$$ structure.  These choices helped us
to initially place and test queries in STS and later adopt them to LTS via
customized python classes.  To clarify QL syntax here is a simple example:
\begin{verbatim}
{"spec": {"task": "/Abc*",
          "timerange":[20160801,20160820]},
 "fields":["error.exitCode"]} ,
\end{verbatim}
which lists a set of conditions defined by a \emph{spec} dictionary, e.g. task
name pattern along with mandatory timerange constraint, and output
\emph{fields} which end-users requested for their result set, e.g. \emph{error.exitCode}
specify a list of error codes to retrive.

It is worth to mention that usage of non-SQL language (JSON) in queries
was easily adopted. Its rich syntax and expresiveness
covers many use cases and usage of JSON in other data-management
applications made this transition effortless.

\subsection{User queries}
Users  place their queries via POST API and fetch the end results via GET API. Here, we provide a
complete example of an end-user interaction with the system via a python client.

A client posts a request which is immediately acknowledged by the system:
\begin{verbatim}
# post request to the server
wma_client.py --spec=query.json
# response from server
{"result": [
    {"status": "ok",
     "input": {"fields": ["wmaid"],
     "spec": {"task": "/Abc*"}},
     "storage": "mongodb",
     "results": [{"wmaid": "6b0bac"}]}]} .
\end{verbatim}
The user query is supplied via the \emph{query.json} JSON file which has the following
content:
\begin{verbatim}
{"spec":{"task": "/Abc*",
         "timerange":[20160801,20160820]},
 "fields":["meta_data"]} .
\end{verbatim}
At a later time the results were accessible via GET API,
e.g. GET \emph{/wmarchive/data/6b0bac} yields the following results:
\begin{verbatim}
{"result": [
    {"status": "ok",
     "input": {"fields": ["meta_data"],
               "spec": {"task": "/Abc*",
      "timerange":[20160801,20160820]}},
     "storage": "mongodb",
     "results": [
     {"meta_data":
        {"agent_ver": "1.0.13.pre8",
         "fwjr_id": "100-0",
         "host": "vocms008.cern.ch",
         "ts": 1454617125}}]}]} .
\end{verbatim}

We didn't provide any specific way to know up-front readiness of the query
and relied on HDFS native solution to look-up job status via its web UI.

\section{Usage of the system}\label{sec:Monitoring}
\begin{center}
\begin{figure}
\includegraphics[width=0.5\textwidth]{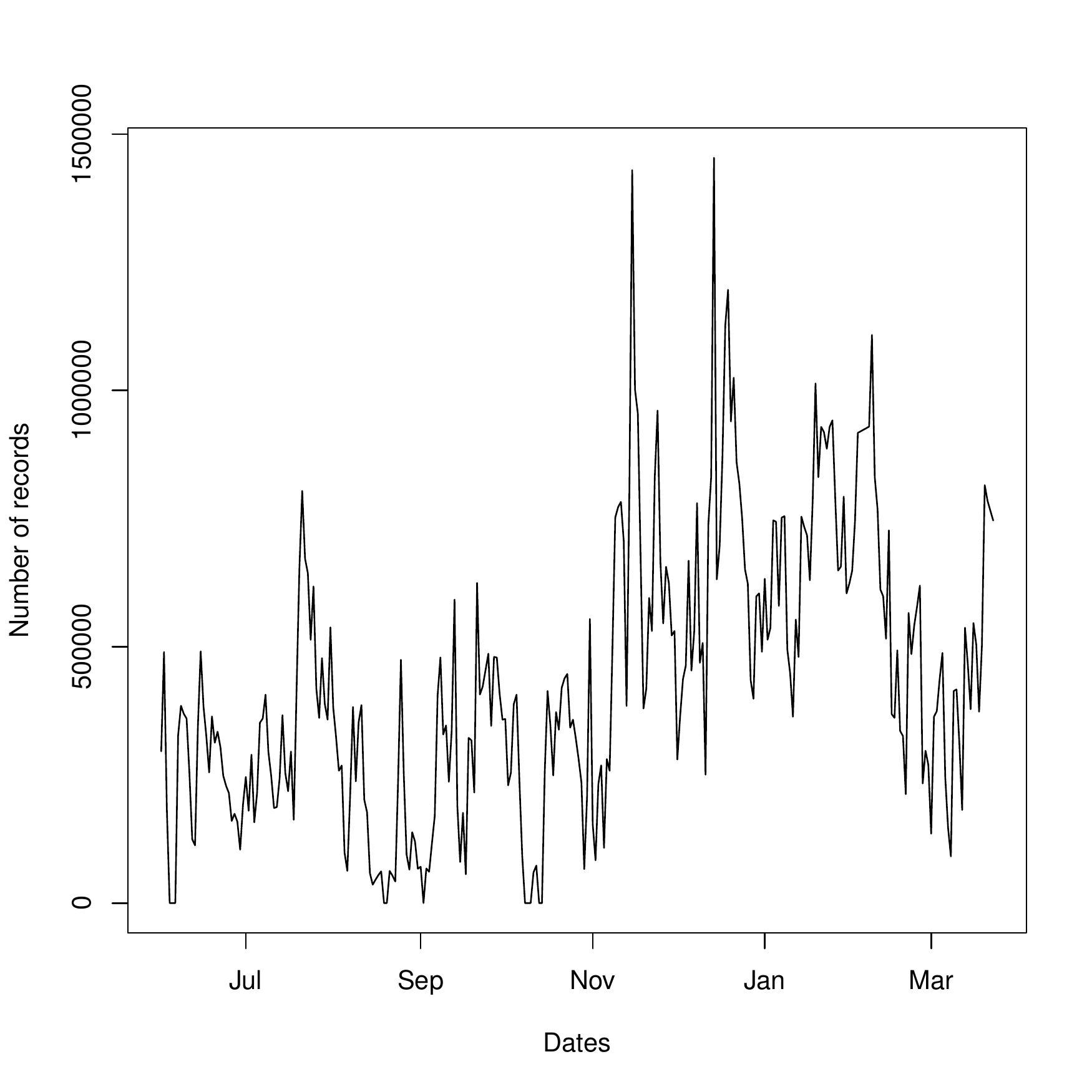}
\caption{Data rate from production CMS WMAgents for a period of July 2016 till March 2017.
    More than 125M documents are collected in WMArchive over this period.}
\label{fig:Rate}
\end{figure}
\end{center}

The WMArchive system was deployed to production in the middle of June 2016.
Since then it has proved to be stable and capable of sustaining the projected
load from the CMS production WMAgents. Fig.~\ref{fig:Rate} shows the data
injection rate for the period from July 2016 until mid March 2017.
We anticipated that load from production agents will eventually grow,
and, as can be seen from Fig \ref{fig:Rate} it was the case. Initially,
we operated at the level of $\mathcal{O}$(200-300K) docs/day and lately we started seeing
5 times higher rate, reaching 1.5M docs/day during busy time of
CMS production. These peaks correspond to a higher data production demand
of CMS collaboration, e.g. the time of the major conferences where end-users
requested more data from data-operations teams to prepare their results
for a conferences.

Since deploying the system into production we monitored its usage and didn't
see significant load on CPU and I/O of the backend node. The RAM usage is
correlated with the STS size (MongoDB) whose data-lookup load depends
on index size of the stored records, see discussion in Sec. \ref{sec:MongoDBIndex}.
Therefore we anticipate that scaling of WMArchive system is more related to Disk and
RAM components rather then CPU and I/O.

The overall cost of the system had minimal impact on CMS budget. We were able
to operate using two CERN VMs, and we relied on centraly supported CERN HDFS
cluster, see Sec \ref{sec:Benchmarks} for more details. The production and
development nodes were configured via puppet and maintained by CMS.

\subsection{Data lookup}\label{sec:DataLookup}
One of the use cases supported by WMArchive is flexible data lookup.
This task is divided into two categories, the data lookup in STS
and in LTS. The former was easy to achive since our QL, see Sec. \ref{sec:QL},
is natively translated into MongoDB queries. The harder part was to
provide ability to look up documents in LTS.
For that we developed custom Python based classes abstracted from a single
interface to follow the map-reduce paradigm. We provide code examples
for common use cases such as couting the number of documents for a given set of
conditions, finding records, finding failed records, and record aggregation.
These classes are fed into a Spark Python wrapper which applied their
logic to a set of documents found on HDFS. This approach helped us
to accommodate more sophisticated tasks required by CMS data-operations teams.
Here, we want to outline the most difficult use case we faced so far.
The task is to find LogCollect files for a given output LFN.


A workflow in an individual CMS WMAgent follows a series of steps where each step
creates its own set of FWJR documents. The normal processing chain contains the
following job sequence:
processing or production job $\rightarrow$ merge job which creates a final set of output files.
The merge jobs are omitted if the processing or production job creates a file bigger
than some threshold. In that case, the output of the production or processing job is the final output file.

For each job, log-archive files are created and the FWJR contains log-archive's
logical file name. Then a LogCollect job creates tar files when enough log-archive
files are collected. The LogCollect job is a separate and independent job which takes
log-archive files as input and produce a tar file as an output.

The common task is to find records from the output file created by merge job (or
processing, production job) and look-up a corresponding tar file created by the
LogCollect job. Also, to find the corresponding log-archive file from processing or
production job.

To accomplish this task we are forced to perform multi-step look-up of
documents stored on HDFS. Here, are the steps we followed:

\begin{itemize}
\item
    look up FWJR documents with provided file name which is present in \emph{LFNArray}
        FWJR list attribute;
\item
    check the job type and, in the case of merge jobs, iterate and retrieve all input files read in the merge job;
\item
    retrieve all unmerged input files from \emph{LFNArray} list;
\item
    for each input file look-up FWJR documents associated with unmerged files in \emph{LFNArray};
\item
    find the file which ends with \emph{logArchive.tar.gz} in \emph{LFNArray} list in the same
        document;
\item
    search again the documents with \emph{LogCollect} job type which contain
        \emph{logArchive.tar.gz} file above as input file;
\item
    return the list of intermediate files along with log-archive and  \emph{LogCollect} job files.
\end{itemize}
This procedure quite often causes three iterations over the data on HDFS to find the
final results. But due to the  excellent parallel processing pipeline on Spark platform
we were able to quickly ($\mathcal{O}(10)$ minutes/week worth of data) find the desired documents on HDFS.

\subsection{Data Aggregation and Monitoring}\label{sec:Aggregation}

Next to long-term storage and flexible access to individual FWJRs, it is one
objective of the WMArchive service to assist CMS data operators in monitoring
the CMS computing infrastructure through an interactive web interface. Of
course,  to access the performance data in the long-term HDFS storage it is
necessary to schedule jobs that retrieve the data, and such tasks can take a
significant amount of time. So to provide a responsive user interface we
constructed an aggregation pipeline that regularly processes the distributed
database of \\
FWJRs to collect performance metrics and cache the aggregated data
in the MongoDB short-term storage, where it is quickly accessible by the
REST server and the user interface.

The primary aggregation procedure based on Apache Spark~\cite{Spark} reduces
the original long-term storage data to a cache of limited size
where only selected information is preserved that data operators may want to commonly
monitor. This cache is not data from each individual job report but is instead
aggregated data grouped only by a number of attributes, for example the job
success state, its host, or its processing site.
During implementation of the aggregation procedure we made several
choices to keep reasonable query times under control.
We took particular care in
selecting a suitable temporal granularity for the aggregation procedure. While
daily aggregated data turned out to be generally sufficient, we keep hourly
aggregated data for a limited time to allow data operators to find failures in
current operations. Stored back in the short-term
storage, a secondary MongoDB aggregation query over the cached
data are produced data for visualizations on timescales suitable for a
responsive user interface.

\begin{figure}
\begin{center}
\includegraphics[width=\linewidth]{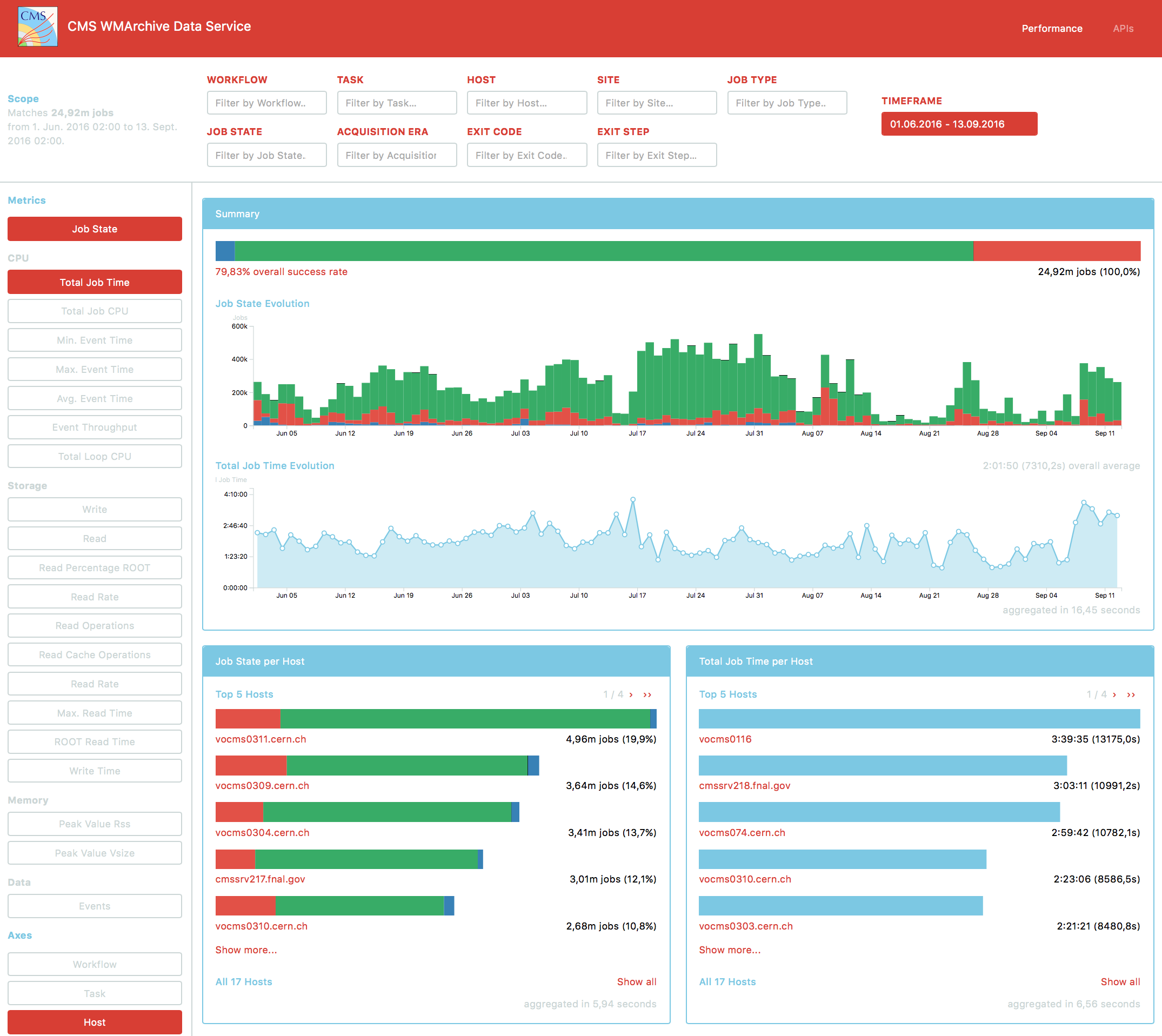}
\caption{The WMArchive Performance Service web interface gives CMS data operators flexible access to a range of commonly needed data aggregation queries.}
\label{fig:WMArchiveWebUI}
\end{center}
\end{figure}

To interactively present the aggregated data, we originally developed the user
interface depicted in Fig. \ref{fig:WMArchiveWebUI} based on open web
technologies such as the D3.js Ja\-va\-Script data visualization
library~\cite{WMArchiveDeps}. We collected feedback from CMS data operators to
construct a user interface that allows them to comfortably perform commonly
needed queries.  They may select any number of performance metrics, such as the
total job computation time, CPU consumption, and storage usage, as well as a
set of visualization axes from the attributes we chose to preserve in the
aggregation procedure.  The service performs the secondary MongoDB aggregation
query for each combination of the selected metrics and axes in realtime and
visualizes the results in a way that is suitable for the query.  The user may
refine the scope of his or her query through regular expression filters by any
of the available attributes, as well as by specifying a timeframe.  Filters are
also applied by interactive user interface elements that allow data operators
to comfortably navigate through the data.

\begin{figure}
\begin{center}
\includegraphics[width=\linewidth]{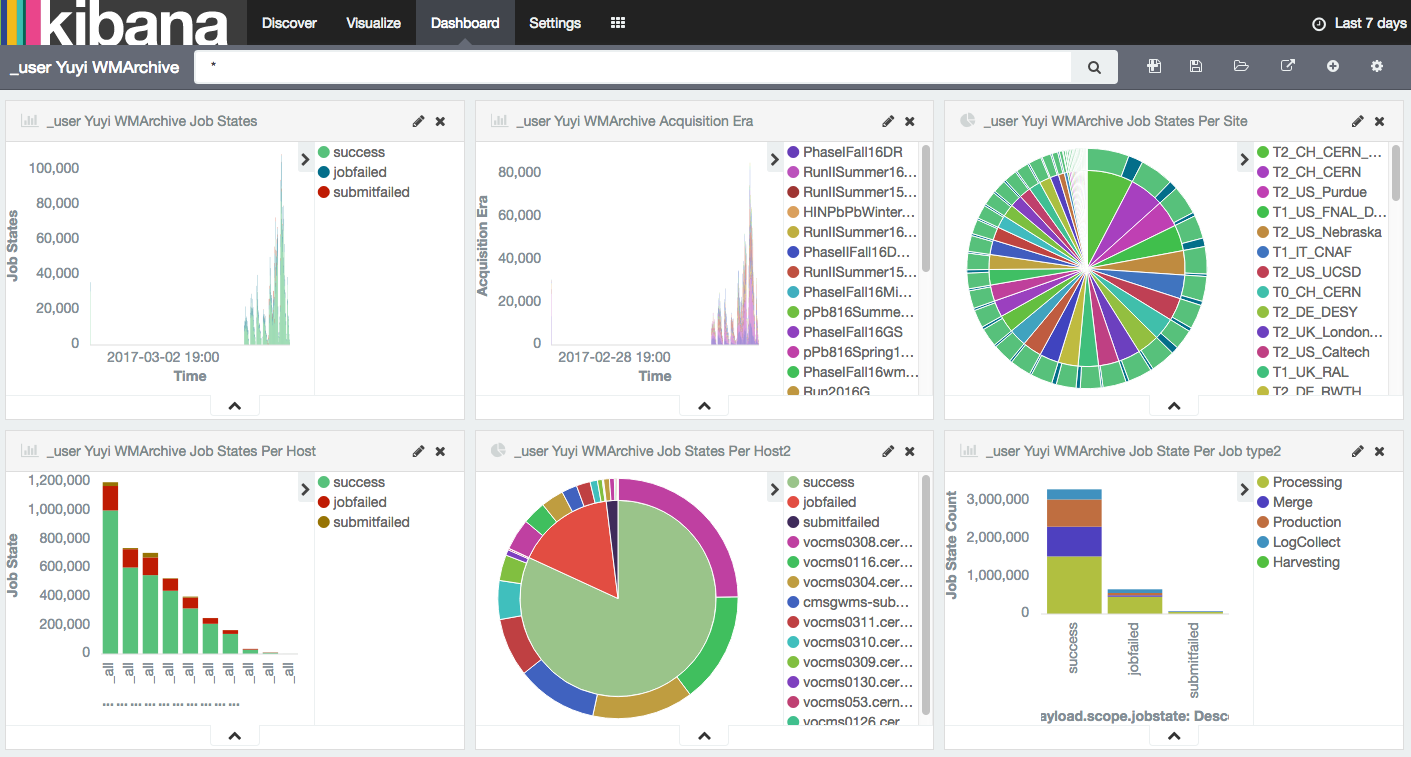}
\caption{The Kibana data visualization platform exposes the entire aggregated WMArchive Performance Service dataset.}
\label{fig:Kibana}
\end{center}
\end{figure}

With the aggregated data available in the short-term storage we also
investigated visualization platforms such as Kibana~\cite{Kibana} and
Grafana~\cite{Grafana} for monitoring the performance data. This service is
provided by the CERN Monitoring system that we feed with the data from the
primary aggregation procedure. Complementing the WMAr\-chive Performance
Service web interface that is tailored to specific workflows of data operators,
these platforms allow for flexible inspection of the entire aggregated dataset
and particularly excel at presenting time series data, as shown in
Fig.~\ref{fig:Kibana}. Although these tools are well adopted and the majority
of CMS data-services use them for central monitoring dashboards we found
that there are certain limitations in their capacity above certain limit,
e.g. slowness of making plots from millions of documents, which are under investigation by CERN IT.

\subsection{Error handling}\label{sec:ErrorHandling}
The error handling is an important component of every system. Apart from usual
programming bugs catched via Exception mechanism in the Python code, we mostly
struggled to handle errors on the Spark side. The PySpark framework is a wrapper
around Java Spark libraries. As such, the nested, often very long and cumbersome,
errors were hard to debug and understand. Most of them came from JVM memory
issues due to incorrect data handling on a worker node. This was caused by a sequential data
model in our code rather than a distributed lazy processing offered by PySpark APIs.
We must admit that a there is learning curve to efficiently use PySpark and adapt it
to the Python code we were dealing with before implementing this system. Due to this nature,
the errors caused by our Spark jobs were separately examined and debugged in collaboration
with CERN IT, while other errors were captured in server logs and invetigated by developers.

\section{Summary}
We have provided a detailed description of the WMArchive system for
distributed workflow management agents in the CMS collaboration.
The system is designed according to a specific set of requirements
imposed by CMS data-operations teams and it has been deployed into
production since mid 2016. Since then we have collected more than
125 million FWJR documents and have not experienced any significant
problems with system maintenance. The introduction of short-term 
and long-term storage systems has helped us to maintain the injection rate and isolate
it from  user-based queries. We found that complex search queries
are desired by the CMS data-operations teams and we are able to accommodate them
via a flexible query language discussed in the paper. We plan to expand
our system and include a new data steam coming out from CMS
CRAB~\cite{CRAB} analysis facilities which process user-analysis jobs across
the globe. This will roughly double the demands on the deployed system where
we may face new challenges. But the flexible design of the system and almost a
year of running in production environment allow us to be optimistic about such
an expansion.

\begin{acknowledgements}
We would like to thank our colleagues Seangchan Ryu (FNAL) and Alan Malta
(Univ. of Nebraska) for numerous feedback and guidance across various
details of CMS Workflow Management System.  Special thanks goes to Eric
Vaandering (FNAL) for initiating the idea of WMArchive system in CMS and
his constant support along development cycle.  We also would like to thank
Luca Menichetti from CERN IT who provided support for development,
maintenance and deployment of our scripts on Spark platform.
\end{acknowledgements}



\end{document}